\documentclass[12pt]{article}
\usepackage{graphicx}
\usepackage{amsfonts}

\begin{document}

\begin{titlepage}

\hfill{Preprint {\bf SB/F/05-336}} \hrule \vskip 2.5cm

\centerline{\bf Zeno and Anti Zeno effect for a two level system
in a squeezed bath }
 \vskip 1.5cm
\centerline{D.F. Mundarain and J. Stephany} \vskip 4mm

\centerline{{\it  Departamento de F\'{\i}sica, Universidad Sim\'on
Bol\'{\i}var,}} \centerline{\it Apartado 89000, Caracas 1080A,
Venezuela.}

\begin{abstract}
We discuss the appearance of  Zeno (QZE) or anti-Zeno (QAE) effect
in an exponentially decaying system. We consider the quantum
dynamics of a continuously monitored two level system interacting
with a squeezed bath. We find that the behavior of the system
depends critically on the way in which the squeezed bath is
prepared. For specific choices of the squeezing phase the system
shows Zeno or anti-Zeno  effect in conditions for which it would
decay exponentially if no measurements were done. This result
allows for a clear interpretation in terms of the equivalent spin
system interacting with a fictitious magnetic field.

\end{abstract}

\vskip 2cm \hrule
\bigskip
\centerline{\bf UNIVERSIDAD SIM{\'O}N BOL{\'I}VAR} \vfill
\end{titlepage}

\section {Introduction}
The suppression or modification of the rate of quantum transitions
in a system, due to successive measurements  is known as the
quantum Zeno effect (QZE) \cite{sud,chiu,peres}. This term has
been applied both to the elimination of the induced transitions as
in the case of Rabi oscillations on a two level system, or to the
reduction of the decay rate on an unstable system. The first
situation was experimentally achieved in 1990 \cite{ita} and the
second one in 2001 \cite{fis}.

An interesting issue in relation with the QZE is wether it appears
or not in exponentially decaying systems. In their article of 1977
Chiu, Sudarshan and Misra \cite{chiu} show that in general, an
unstable system has three decaying regimens. For short time
intervals, $t \leq T_1$  or very large ones $t\geq T_2$,  with
$T_1$ and $T_2$ some time scales, the system depart from the
exponentially decaying behavior shown for $T_1 \leq t \leq T_2$.
They also predict that frequent measurements led to the QZE  if
the time interval between successive measurements is shorter than
$T_1$. In the experiment of Ref. \cite{fis} for example, for short
times, the decay rate of the system is remarkably slower than
exponential. This could lead to think that QZE only occurs when
the time between measurements is short enough to exploit the
departure from the exponential decay law. Nevertheless in a recent
article Koshino and Shimizu \cite{kos1} predicted the appearance
of QZE even for systems with an exponential decay law in the case
when the detector has a finite window of sensibility. For this
case they analyze explicitly  the interaction between the quantum
system and the detector and interpreted the changes induced by the
interaction as the effect of the measurement. They refer
\cite{kos2} to  this analysis as the dynamical formalism as
opposed to the conventional formalism where the measurements are
taken as projections consistent with  the quantum collapse
postulate of von Neumann.

For a closed system the theoretical description of the measurement
in terms of the projection postulate predicts a complete  Zeno
effect, that is the freezing of the quantum system in the initial
state. For such system with a hamiltonian  ${\it H}$, the
evolution is determined by the Schr{\"o}dinger equation
\begin{equation}
\frac{\partial}{\partial t} |\psi(t) \rangle = \frac{1}{i \hbar}
{\it H} |\psi(t) \rangle
\end{equation}

If the observable ${\it A}$ to be measured has eigenvalues $a_m$
and supposing  that at $t=0$ the system is in the eigenstate
$|a_n\rangle$, the probability of obtaining the result $a_n$ for a
short time interval  $\Delta t\geq 0$ is given by,
\begin{equation}
P_n(\Delta t) = \left( 1 - \frac{\Delta t^2}{\hbar^2} \Delta_n^2
{\it H}\right)
\end{equation}
where
\begin{equation}
\Delta_n^2 {\it H} = \langle a_n |{\it H}^2|a_n\rangle -\langle
a_n |{\it H}|a_n\rangle^2
\end{equation}
If one considers   $S$ successive projective measurements
separated by the same interval $\Delta t$ the probability of
obtaining in each case the same result $a_n$ is:
\begin{equation}
P_n(S,\Delta t) = \left( 1 - \frac{\Delta t^2}{\hbar^2} \Delta_n^2
{\it H}\right)^S
\end{equation}
In the limit of very  frequent measurements \cite{Braginsky}, that
is when $ S >> 1 , \ \ \ \ S \Delta t  \rightarrow t$ the
probability of measuring $a_n$ every time is
\begin{equation}
P_n^{(cm)} (t) = \lim_{S \rightarrow  \infty} P_n(S, t/ S) =
\lim_{S \rightarrow  \infty}\left( 1 - \frac{t^2}{S^2 \hbar^2}
\Delta_n^2 {\it H}\right)^S = 1
\end{equation}
which corresponds to a complete Zeno effect.

For open systems in contact with the environment some limitations
affect the appearance of the QZE even if projective measurements
are being done. For  time intervals which are  greater than the
correlation time of the bath, the evolution may  be described in
terms of the density matrix by a master equation of the Liouville
type,
\begin{equation}
\label{Liuv} \frac{\partial\rho}{\partial t} = {\it L } \left\{\rho\right\}  .
\end{equation}
with  ${\it L } \left\{\rho\right\}$ some appropriate operator
depending on $\rho$. Then, for a short time interval   $\Delta t$,
the density operator is given in terms of its initial value by
\begin{equation}
\rho( \Delta t ) = \rho(0) + {\it L}\left\{ \rho(0)\right\} \Delta t
\end{equation}
If the initial state is  $\rho(0) = |a_n\rangle \langle a_n| $ the
probability of measuring $a_n$ in $S$ consecutive measurements
separated by  time intervals $\Delta t$ is,
\begin{equation}
P_n(S,\Delta t)  =( 1 + \langle a_n|{\it L}\left\{
\rho(0)\right\}|a_n\rangle \Delta t)^S\simeq (\exp \left\{\langle
a_n|{\it L}\left\{ \rho(0)\right\}|a_n\rangle\ \Delta t\right\})^S
\end{equation}
In the limit $ S >> 1 , \ \ \ \ S \Delta t  \rightarrow t$, of
very frequent measurements  one obtains,
\begin{equation}\label{ec13}
P_n^{(cm)}(t)= \exp \left\{\langle a_n|{\it L}\left\{ \rho(0)\right\}|a_n\rangle\ \
t\right\}
\end{equation}
Here the freezing of the initial condition for continuous
measurements is achieved only if
\begin{displaymath}
 \langle a_n|{\it L}\left\{\rho(0)\right\}|a_n\rangle  = 0
\end{displaymath}
This illustrates the fact that in general, both the the
intrinsical properties of the system and the characteristics of
the measurement affect the possibility of  displaying the quantum
Zeno effect.

A related issue that we have to consider comes from the
observation that for an unstable quantum system  the probability
of obtaining a specific result in a measurement  may   increase,
decrease or even oscillate in time as the result of its
undisturbed evolution. Decay rates may also be affected by
measurements done at particular instants of time, an effect which
has in principle nothing to do with the QZE. This  suggests that
the interaction of the system with a non trivial electromagnetic
bath may  modify the decay rates  even for an exponentially
decaying system. In this paper we show that such mechanism can be
actually used to induce QZE or QAE in a two level system. For this
system interacting with a squeezed bath QZE or QAE may appear when
measuring the fictitious spin along a specific direction depending
on the relative phase of the squeezing and the chosen direction.
This may be interpreted as an effect of the orientation induced on
the fictitious spin by the fictitious magnetic field defined by
the quadratic fluctuations of the true electromagnetic field.

\section{The two level system in a squeezed bath}
In the rotating wave approximation the hamiltonian which better
describes the atom-field  interaction has the following structure,
\cite{gar,scully}:
\begin{equation}
{\it H} = \sum_{\bf k} \hbar \nu_k a_{\bf k}^{\dagger} a_{\bf k} +
\frac{1}{2} \hbar \omega \sigma_z+ \hbar \sum_{\bf k} g_{\bf k}
\left( \sigma_{+} a_{\bf k}+a_{\bf k}^{\dagger} \sigma_{-} \right)
\label{hamil}
\end{equation}
where $g_{\bf k}$ are the atom-field couplings constants, $ a_{\bf
k}$ and  $a_{\bf k}^{\dagger}$ are the creation and annihilation
operators of the multimodal field and  $\sigma_+$ and $\sigma_-$
are the ladder operators
\begin{equation}
\sigma_{+} =\left(
\begin{array}{cc}
0&1\\
0&0
\end{array}
\right)\qquad \sigma_{-} =\left(
\begin{array}{cc}
0&0\\
1&0
\end{array}
\right)\ \  ,
\end{equation}
with $\sigma_x$, $\sigma_y$ and $\sigma_z$ are the Pauli matrices,
\begin{equation}
\sigma_x =\left(
\begin{array}{cc}
0&1\\
1&0
\end{array}
\right)\qquad \sigma_y =\left(
\begin{array}{cc}
0&-i\\
i&0
\end{array}
\right) \qquad  \sigma_z =\left(
\begin{array}{cc}
1&0\\
0&-1
\end{array}
\right)\ \ .
\end{equation}
If the field is prepared in a broadband squeezed vacuum state
characterized by $\xi=r e^{i\phi}$ it was demonstrated that,
\cite{gar,scully}:
\begin{displaymath}
\langle a_{\bf k}\rangle =\langle a_{\bf k}^{\dagger}\rangle=0
\end{displaymath}
\begin{displaymath}
\langle a_{\bf k}^{\dagger} a_{\bf k'}\rangle \delta_{\bf k k'} =
N \delta_{\bf k k'}
\end{displaymath}
\begin{displaymath}
\langle a_{\bf k} a_{\bf k'}^{\dagger}\rangle= \delta_ {\bf k k'}=
(N+1) \delta_{\bf k k'}
\end{displaymath}
\begin{displaymath}
\langle a_{\bf k} a_{\bf k'} \rangle= - e^{i \phi} \cosh ( r)
\sinh(r) \delta_{\bf k', 2 k_0 - k} = e^{i \phi} M \delta_{\bf k',
2 k_0 - k}
\end{displaymath}
\begin{equation}\label{ec4}
\langle a_{\bf k}^{\dagger} a_{\bf k'}^{\dagger} \rangle=-  e^{-i
\phi} \cosh ( r) \sinh(r) \delta_{\bf k', 2 k_0 - k} = e^{-i \phi}
M \delta_{\bf k', 2 k_0 - k}\ \ ,
\end{equation}

\noindent where $N= \sinh^2 ( r)$, $M=\sqrt{N(N+1)}$. Here ${\bf
k_0}$ is the wave number associated to the resonant frequency of
the squeezing device. In the interaction picture the master
equation for this system takes the form of Eq. (\ref{Liuv}) with,
\begin{eqnarray}\label{em1}
 L\{\rho\}  =&\frac{1}{2}\gamma \left( N+1\right) \left( 2 \sigma_{-}
{ \rho} \sigma_{+} - \sigma_{+} \sigma_{-}  { \rho} - {\rho}
\sigma_{+} \sigma_{-}
\right)\nonumber\\
&\frac{1}{2} \gamma N \left( 2 \sigma_{+}  { \rho} \sigma_{-} -
\sigma_{-} \sigma_{+}  {\rho} - { \rho} \sigma_{-} \sigma_{+} \right)\nonumber\\
&- \gamma  M e^{ i \phi}  \sigma_{+}  { \rho} \sigma_{+} -\gamma M
e^{ -i \phi}  \sigma_{-}  { \rho} \sigma_{-}\ \ .
\end{eqnarray}
Here $\gamma$ is the decay constant of the system in the vacuum.
This equation may be rewritten using Bloch's representation for
the two level system density matrix in the form,
\begin{equation}\label{dm1}
\rho = \frac{1}{2} \left( 1 + \rho_{x} \sigma_{x}+\rho_{y}
\sigma_{y} + \rho_{z} \sigma_{z} \right)\ \ .
\end{equation}

\begin{figure}[ht]
\includegraphics[scale=0.7]{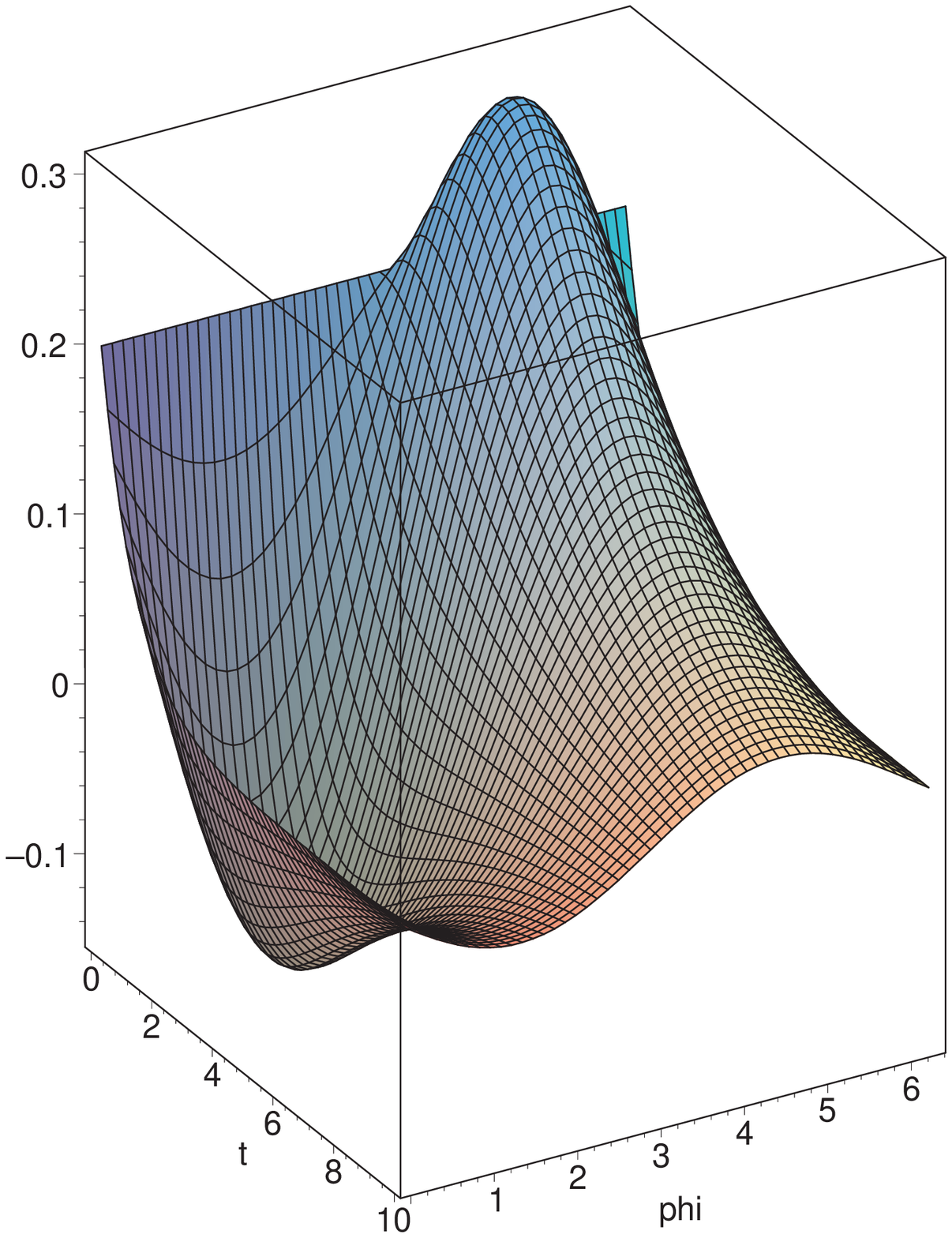}
\caption{$\rho_x(t,\phi)$, $\rho_x(0)=0.2$ ,
$\rho_y(0)=-\sqrt{1-\rho_x(0)^2}$,  $\rho_z(0)=0$, $N=1$, $\gamma =1$}\label{fig1}
\end{figure}

Using Eqs. (\ref{em1},\ref{dm1}), the master equation (\ref{Liuv})
takes the form,
\begin{eqnarray}
\frac{\partial \rho}{\partial t}  =& -\frac{1}{2} \gamma \left(
N+1\right) \left( (1+\rho_z) \sigma_{z} +\frac{1}{2}
\rho_x \sigma_{x}  +\frac{1}{2} {\rho_y} \sigma_{y} \right)\nonumber\\
& +\frac{1}{2} \gamma N \left( (1-\rho_z) \sigma_{z} -\frac{1}{2}
\rho_x \sigma_{x}  -\frac{1}{2} {\rho_y} \sigma_{y} \right)\nonumber\\
& -\frac{1}{2} \gamma M  \rho_{x} ( \cos(\phi)\sigma_{x} -\sin
(\phi) \sigma_{y}) \nonumber\\
&+\frac{1}{2} \gamma M  \rho_{y} ( \sin(\phi)\sigma_{x} +\cos
(\phi) \sigma_{y})
\end{eqnarray}
This is equivalent to the following  differential equations for
$(\rho_x, \rho_y, \rho_z)$:
\begin{displaymath}
\dot{\rho_x} = -\gamma \left(N+1/2+M \cos (\phi)\right) \rho_x +
\gamma M \sin(\phi) \rho_y
\end{displaymath}
\begin{displaymath}
\dot{\rho_y} = -\gamma \left(N+1/2-M \cos (\phi)\right) \rho_y +
\gamma M \sin(\phi) \rho_x
\end{displaymath}
\begin{equation}
\dot{\rho_z} = -\gamma \left(2N+1/2 \right) \rho_z -\gamma
\end{equation}
The solutions of these equations are given by,

\begin{eqnarray}\label{ec1222}
\rho_x(t)  =& \left( \rho_x (0) \sin^2(\phi/2) +\rho_y(0)
\sin(\phi/2) \cos( \phi/2) \right) e^{ -\gamma (N+1/2-M)\, t}
\nonumber\\
&+\left( \rho_x (0) \cos^2(\phi/2) -\rho_y(0) \sin(\phi/2) \cos(
\phi/2) \right) e^{ -\gamma (N+1/2+M)\, t}
\end{eqnarray}
\begin{eqnarray}
\rho_y(t)  =& \left( \rho_y (0) \cos^2(\phi/2) +\rho_x(0)
\sin(\phi/2) \cos( \phi/2) \right) e^{ -\gamma (N+1/2-M)\, t}
\nonumber\\
&+\left( \rho_y (0) \sin^2(\phi/2) -\rho_x(0) \sin(\phi/2) \cos(
\phi/2) \right) e^{ -\gamma (N+1/2+M)\, t}
\end{eqnarray}

\begin{equation}
\rho_z(t) = \rho_z(0) e^{-\gamma (2 N+1) t} + \frac{1}{2 N +1}
\left(  e^{-\gamma (2 N+1) t}-1\right)
\end{equation}
From these expressions one can read the dependence of the decay
rates of the system on the phase $\phi$ of the squeezing. In
particular, for $\phi=0$, $\phi=\pi$ or for the  critical angles $
\phi_z = 2 \arctan(-\rho_y(0) / \rho_x(0))$ or $\phi_{AZ}= 2
\arctan (\rho_x(0)/ \rho_y(0) )$, the system has a purely
exponential behavior with the decay rates presented in Table \ref{Table1}.
\vskip.3cm
\begin{table}[t]
\begin{center}
\mbox{\begin{tabular}{|r|r|r|} \hline
\ & $\rho_x (t)/\rho_x(0)$ & $\rho_y (t)/\rho_y(0)$ \\
\hline
$\phi=0\quad$ & $e^{-\gamma (N+1/2+M) t }$&  $e^{-\gamma (N+1/2-M) t }$ \\
$\phi=\pi \quad$ &  $e^{-\gamma (N+1/2-M) t}$ & $e^{-\gamma (N+1/2+M) t }$ \\
$\phi=\arctan \left(-\frac{\rho_y(0)}{\rho_x(0)}\right)$ &
$e^{-\gamma (N+1/2+M) t}$ & $e^{-\gamma (N+1/2+M)t}$  \\
$\phi=\arctan \left(\frac{\rho_x(0)}{\rho_y(0)}\right)$ &
$e^{-\gamma (N+1/2-M) t}$ & $e^{-\gamma (N+1/2-M)t}$ \\ \hline
\end{tabular}}
\end {center}
\caption{Decay rates for critical angles}
\label{Table1}
\end{table}
In Fig.(\ref{fig1}) we show the dependence of $\rho_x(t) = \langle
\sigma_x \rangle$ with the phase as given by Eq. (\ref{ec1222}).
In particular the exponential decay for the preferred values of
the phase may be observed.

\section{The origin of the critical angles} Before discussing the effect of the
measurements in the evolution of the two level system let us first
explore the properties of the fictitious magnetic field associated
to the squeezed state in order to justify  the decay  rates for
the two critical angles appearing in Table 1.

Consider the atomic part of the Hamiltonian (\ref{hamil}). In
terms of the Pauli matrices it takes the form,
\begin{equation}
{\it H}_{Atomic} =  \frac{1}{2} \hbar \omega \sigma_z+ \frac{1}{2}
\hbar \sigma_x \sum_{\bf k} g_{\bf k} \left( a_{\bf k}+a_{\bf
k}^{\dagger}\right) + \frac{1}{2} \hbar \sigma_y \sum_{\bf k} i
g_{\bf k} \left( a_{\bf k}- a_{\bf k}^{\dagger}\right)\ .
\end{equation}
This can be  rewritten in the form
\begin{equation}
{\it H}_{Atomic} =  - \gamma_0 \,  {\bf B \cdot S}\ \ .
\end{equation}
where $\gamma_0$ is an arbitrary constant with dimensions of
charge divided by  mass, ${\bf S}$ is the fictitious spin
associated to the two level system and  ${\bf B}$ is the quantum
fictitious magnetic field with components,
\begin{equation}
B_x = -\frac{1}{\gamma_0} \sum_{\bf k} g_{\bf k} \left( a_{\bf
k}+a_{\bf k}^{\dagger}\right)
\end{equation}
\begin{equation}
B_y = - \frac{1}{\gamma_0} \sum_{\bf k} i  g_{\bf k} \left( a_{\bf
k}- a_{\bf k}^{\dagger}\right)
\end{equation}
\begin{equation}
B_z = - \frac{\omega}{\gamma_0}\ \ .
\end{equation}
Clearly  $<B_x>=0$ and $<B_y>=0$. For the quadratic fluctuations
the result is,
\begin{eqnarray}
<B_x^2>&=&\frac{\Gamma}{4}\left(\sinh^2(r)+\cosh^2(r)-2\cos(\theta)\sinh(r)\cosh(r)\right)\\
&=&\Gamma \left( N +\frac{1}{2}- M \cos(\phi) \right)\nonumber\\
<B_y^2>&=&\frac{\Gamma}{4}\left(\sinh^2(r)+\cosh^2(r)+2\cos(\theta)\sinh(r)\cosh(r)\right)\\
&=& \Gamma \left( N +\frac{1}{2}+ M \cos(\phi) \right)\nonumber
\end{eqnarray}
where
\begin{displaymath}
\Gamma = \frac{1}{2 \gamma_0^2} \sum_{\bf k}| g_{\bf k}|^2 .
\end{displaymath}
Here $\Gamma=\sum_k |g_k|^2$ is taken to be finite, which means
that only a finite subset of the modes in the bath is coupled
effectively to the system. For $<B^2>$ we have,
\begin{eqnarray}
<B^2>=\frac{\Gamma}{2}\left(\sinh^2(r)+\cosh^2(r)\right) =
\frac{\Gamma}{2}\left( 2 N +1 \right) ,
\end{eqnarray}
which does not depend on $\phi$.

\begin{figure}[ht]
\centerline{\includegraphics[scale=0.7]{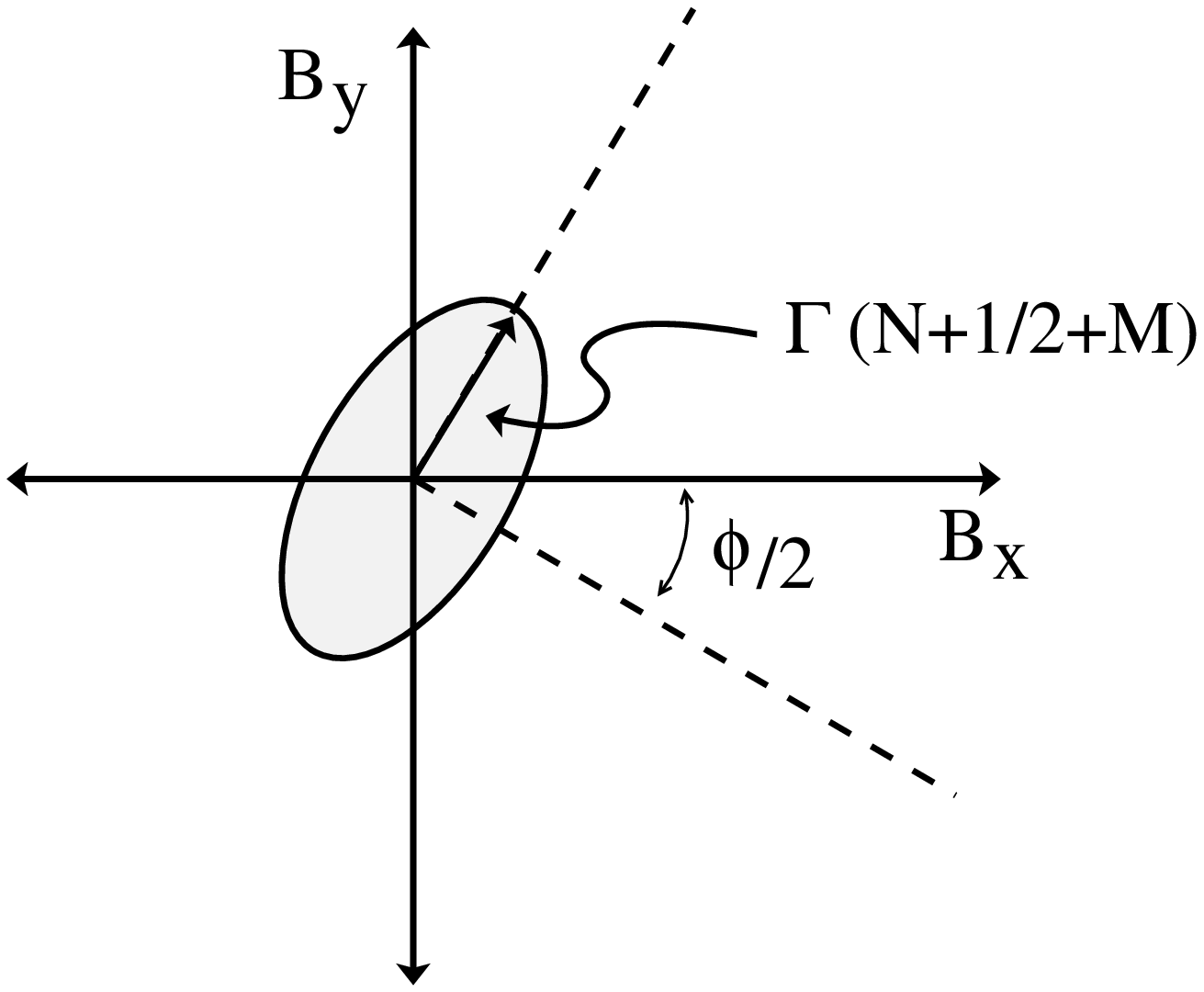}}
\caption{Fluctuations of the fictitious magnetic field.}
\label{fig17}
\end{figure}

These fluctuations may be represented in phase space as an ellipse
whose axis are rotated by an angle  $\phi/2$.  As is illustrated
in Fig. (\ref{fig17}), the semi-axis have magnitudes   $\Gamma
\left( N +1/2+ M \right)$ and $\Gamma \left( N +1/2- M \right)$.

 Comparing with the results of  the previous section, we can observe,
that for zero phase  the decay rate for $\rho_x = \langle
\sigma_x\rangle$ is proportional to the fluctuations of the
fictitious magnetic field component $B_y$ and the decay  rate for
$\rho_y = \langle \sigma_y\rangle$ is proportional to the
fluctuations of $B_x$. Also,  the  decay rate   for $\rho_z$ is
proportional to $<B^2>$.  In general, for other values of the
phase, the component $(\rho_x, \rho_y)$ of Bloch's vector
orthogonal to the major semi-axis of the phase space ellipse used
to represent the magnetic field fluctuations, has a decay rate
proportional to $\Gamma \left( N +1/2+ M \right)$ and the
component orthogonal to the minor semi-axis has a decay rate
proportional to $\Gamma \left( N +1/2- M \right)$. The phase
$\phi_Z=2 \arctan \left(-\rho_y(0)/ \rho_x(0)\right)$ defines a
critical value which corresponds to the case when initially the
Bloch vector is orthogonal to the major semi-axis. For this value
$\rho_x$ and $ \rho_y$ decay with the maximum rate. The
complementary case occurs for $\phi_{AZ}= 2 \arctan ( \rho_x(0)/
\rho_y(0))$ in which case $\rho_x$ and $ \rho_y$ decay with the
minimum allowed value of the decay rate.

The fact that the decay rates for $\rho_x$ and $ \rho_y$ coincides
in both cases  is a consequence of the coupled dynamics of these
two variables. But if one measures  $\sigma_x$, the dynamics
disentangles  and one would expect that the decay rate for $
\rho_x$ results  proportional  to  the fluctuations of $B_y$ only
, as in the cases $\phi=0$ or $\phi=\pi$ when  there is no
coupling at all. Then, we expect the Zeno effect to occur  for
$\phi_Z$ and the anti-Zeno effect to occur for $\phi_{AZ}$  due to
the factor $\cos (\phi)$ that appears  in the fluctuation of
$B_y$.
\section{Zeno and anti-Zeno effect}

Let us now consider explicitly  the effect of repeated
measurements of the observable  $\sigma_x$ in the  evolution of
the system prepared in a state defined by the initial values of
$(\rho_x(0),\rho_y(0), \rho_z(0))$. We  suppose that the time
interval between measurements is very short, but still much
greater than the correlation time of the squeezed bath
\cite{GarPC1987}. Then we may describe the evolution of the system
by means of a master equation of the form (\ref{Liuv}). In our
analysis we take in fact the  correlation time of the squeezed
bath to be zero which corresponds to broadband squeezing. For
considerations on the finite bandwidth effects se Ref.
\cite{GarPC1987,ParG1988,TanE1998}. On the experimental side
squeezing with a bandwidth of up to 1GHz has been reported
\cite{Cro1988,MacY1988,HirKI2005}.

The probability that in a very large succession of measurements,
the result obtained in all of them is the eigenvalue $+1$
associated to the eigenstate $|+\rangle_x$ is given by,
\begin{equation}
\label{ec217} P_{+}^{(cm)}(t) = \frac{(1+\rho_x(0))}{2} \exp
\left\{ _x \langle+|{\it L}\left\{ \rho_1 \right\}|+\rangle_x t
\right\}
\end{equation}
where $\rho_1$ is the collapsed density matrix after the
measurements and is given by
\begin{equation}
\rho_1=|+\rangle_{x\, x} \langle +| \ \ \ , \ \ \ |+\rangle_x =
\frac{1}{\sqrt{2}}\left( |+\rangle + |-\rangle \right) .
\end{equation}
The probability of Eq. (\ref{ec217}) is obtained by multiplying
the probability corresponding to the first measurement and the
probability obtained in Eq. (\ref{ec13}) which is valid for the
following measurements.  If the system is initially in the state
$|+\rangle_x$ then $\rho_x(0) =1$, $\rho_y(0) =0$ and $
\rho_z(0)=0$, in which case Eq. (\ref{ec217}) is a particular case
of Eq. (\ref{ec13}).

One can show that for the squeezed bath,
\begin{equation}
 _x \langle+|{\it L}\left\{  \rho_1\right\}|
 +\rangle_x =- \frac{\gamma}{2}\left( N +\frac{1}{2} + M \cos (\phi)
 \right).
\end{equation}
In this case Eq. (\ref{ec217}) reduces to,
\begin{equation}\label{ec218}
P_{+}^{(cm)}(t,\phi) =\frac{(1+ \rho_x(0))}{2} \exp \left\{ -
\frac{\gamma}{2}\left( N +\frac{1}{2} + M \cos (\phi) \right) t
\right\}
\end{equation}
We should compare this expression with the probability of
measuring the eigenvalue $+ 1$ by performing an unique measurement
at time $ t$
\begin{equation}\label{ec33}
P_+(t,\phi) =\frac{(1+\rho_x(t))}{2} .
\end{equation}

\begin{figure}[ht]
\includegraphics[scale=0.6,angle=-90]{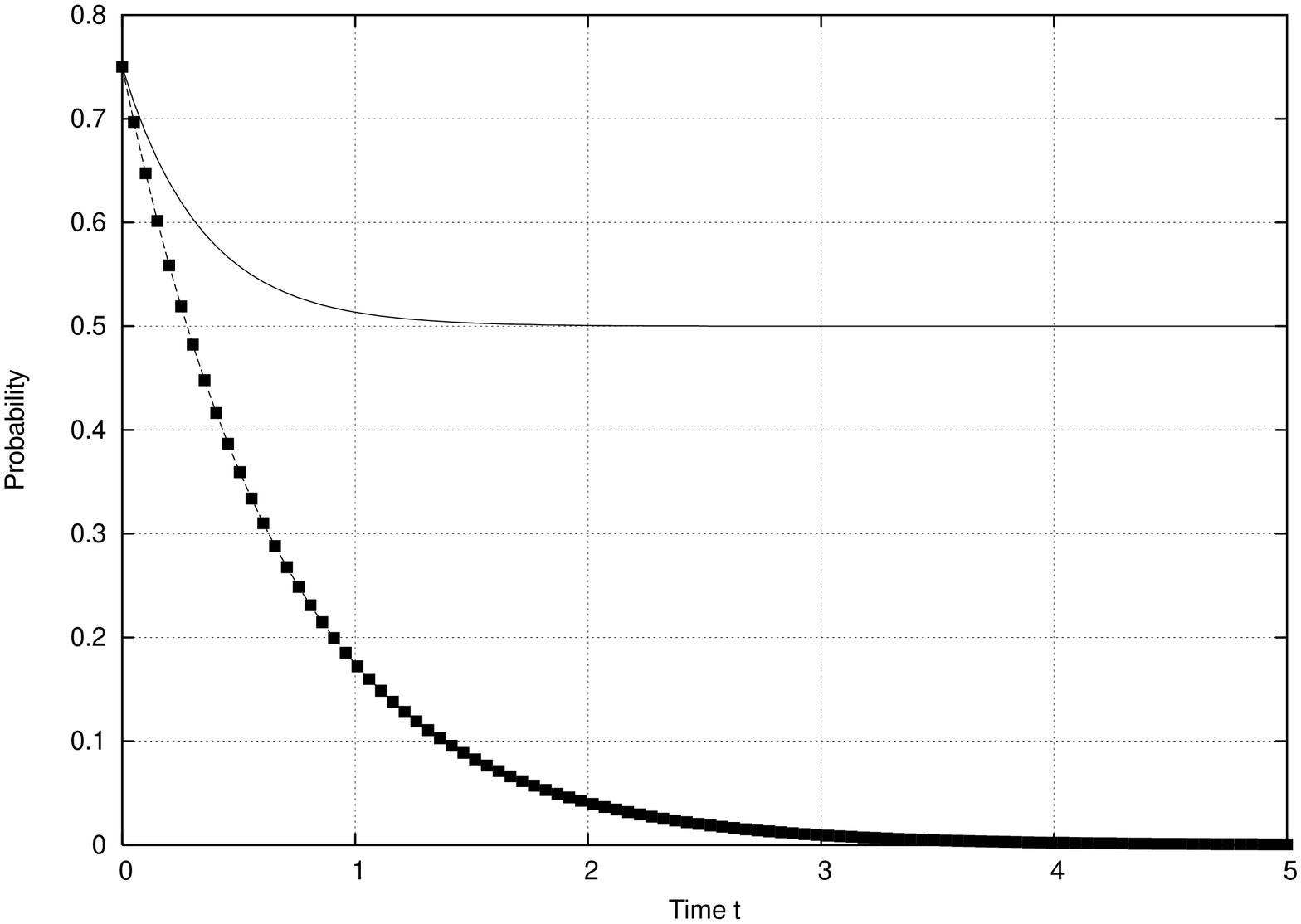}
\caption{Solid line: $P_+ (t)$. Dashed line: $P_+^{(cm)}
(t,\phi=0)$. $\rho_x(0)=0.5$, $\rho_y(0)=-\sqrt{1-\rho_x(0)^2}$, $\rho_z(0)=0$,
$N=1$, $\gamma=1$ }\label{fig2}
\end{figure}

In Fig.(\ref{fig2}) we show that the probability $P_+ (t,\phi=0)$
decays exponentially to the value $1/2$, that is,  for $t
\rightarrow \infty$ we have the same probability to measure any of
the two eigenvalues. On the other hand the probability $P_+^{(cm)}
(t,\phi=0)$, i.e the probability to obtain the same value $+1$ in
all the measurements,  decays exponentially to zero. In the same
figure we can see that the probability $P_+^{(cm)} (t,\phi=0)$ is
smaller than the probability $P_+ (t,\phi=0)$ for all $t$. Note
that, the probability of obtaining the result $+1$ in the last
measurement independently of the previous results is of course
greater than the probability of obtaining the value $+1$ in all
the measurements. Furthermore, if the evolution of the observable
$\sigma_x$ is not affected by the measurements, the probability to
obtain $+1$ in the last measurement independently of the results
of the previous measurements is equal to the probability to obtain
$+1$ at time $t$ if no other measurement has been done previously.
The result shown in Fig.(\ref{fig2}) suggest that in fact for
$\phi=0$ the evolution of the observable $\sigma_x$ is not
affected by the measurements.

\begin{figure}[ht]
\includegraphics[scale=0.6,angle=-90]{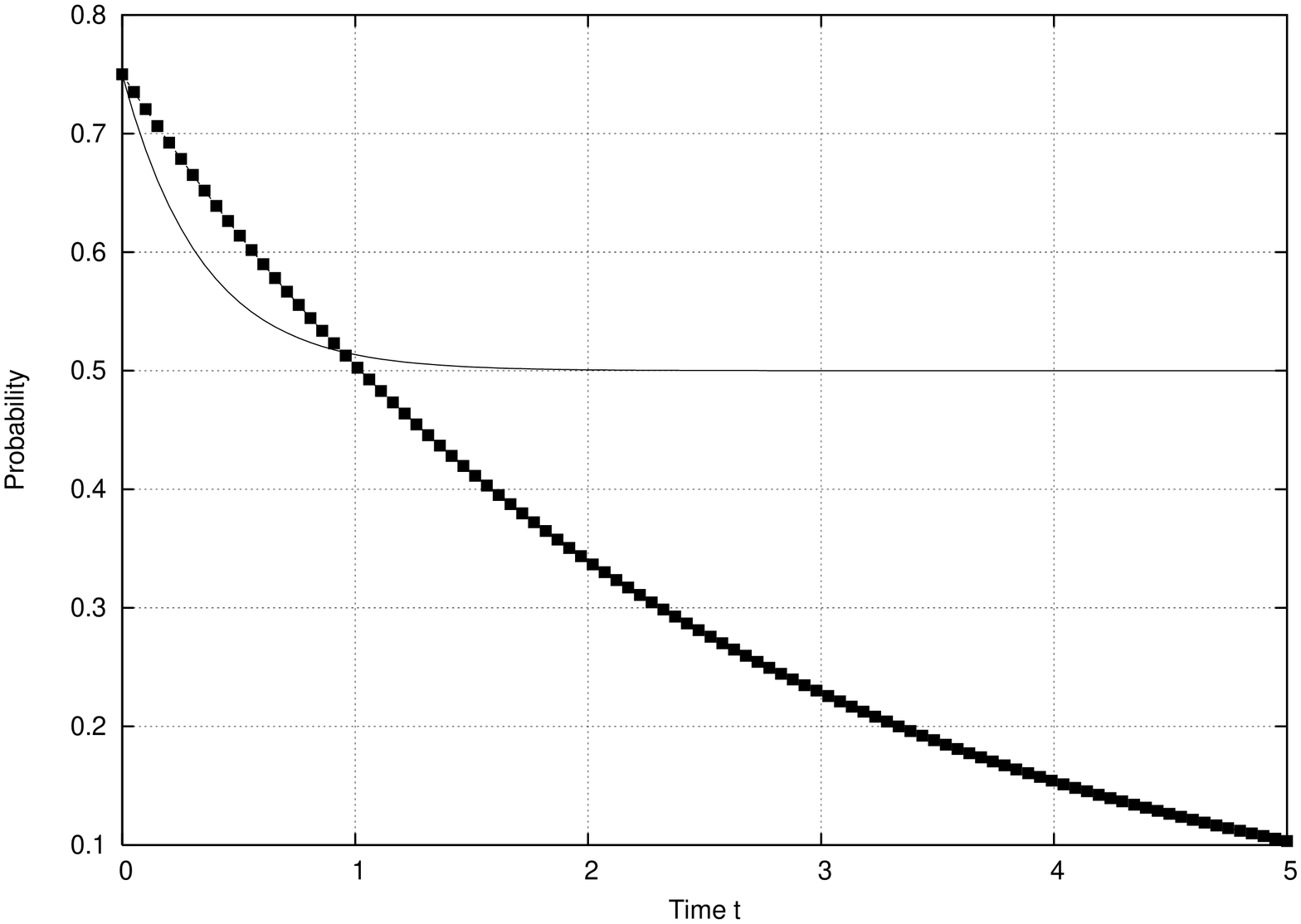}
\caption{Solid line:  $P_+ (t)$. Dashed line: $P_+^{(cm)}
(t,\phi_Z)$. $\rho_x(0) =0.5$, $\rho_y(0)=-\sqrt{1-\rho_x(0)^2}$, $\rho_z(0)=0$,
 $N=1$,  $\gamma=1$
}\label{fig3}
\end{figure}

Changing the phase it  is possible to obtain a completely
different result.  In Fig (\ref{fig3}) we show that  there exists
a time interval for which $P_+^{(cm)} (t,\phi_Z)$ is greater than
the $P_+ (t,\phi_Z)$. The natural explanation for this, comes from
the fact that in this case the measurements do modify the dynamics
of the observable.

To study quantitatively this effect it is necessary to work out
the changes in the master equation related to the continuous
monitoring. If we have the system described by $\rho$ and perform
measurements of $\sigma_x$  the new density matrix is given by,
\begin{equation}
\rho^{\prime} = P \rho P + (1-P)\rho (1-P)
\end{equation}
where  $P=|+\rangle_{xx}\langle+|$ is the projector to the
eigenvector of $\sigma_x$ with eigenvalue $+ 1$ and $(1-P)$ is the
projector to the eigenvector of $\sigma_x$ with eigenvalue $- 1$.

Between consecutive  measurements the free evolution is determined
by the free master equation. By considering the free master
equation and  the collapse in the same  expression, it is shown
that after the first measurement the master equation with
continuous measurements takes the form,
\begin{equation}
\frac{\partial \rho}{\partial t} = P {\it L}(\rho)P  +(1- P) {\it
L}(\rho)(1-P)
\end{equation}

Let us now focus in the mean value of the measured observable
\mbox{$\langle \sigma_x\rangle = \rho_x(t)$}.  The corresponding
probabilities may be computed using Eq. (\ref{ec33}). In terms of
Bloch's vector, the master equation with continuous measurements
for the two level system in the squeezed bath is given by the
following  equations,
\begin{eqnarray}
\dot{\rho_x}& = & -\gamma( N+1/2+M \cos (\phi) )\rho_x+ \gamma M \sin (\phi) \rho_y\nonumber\\
\dot{\rho_y}& = &0\nonumber\\
\dot{\rho_z}& = &0\label{ec39}
\end{eqnarray}
Since after the first measurement the values of $ \rho_y$ and
$\rho_z$ collapse to zero, the solutions for this system are given
by ,
\begin{equation}\label{ec332}
\rho_x(t) = \rho_x(0) \exp ( -\gamma( N+1/2+M \cos (\phi) )t)
\end{equation}

\begin{equation}
\rho_y(t) = 0 \qquad \rho_z(t) =0    .
\end{equation}
As we can see from Eq. (\ref{ec332}) in presence of very frequent
measurements the decay  rate  of $\rho_x$ is proportional to the
quantum fluctuation of $B_y$.

\begin{figure}[ht]
\includegraphics[scale=0.6,angle=-90]{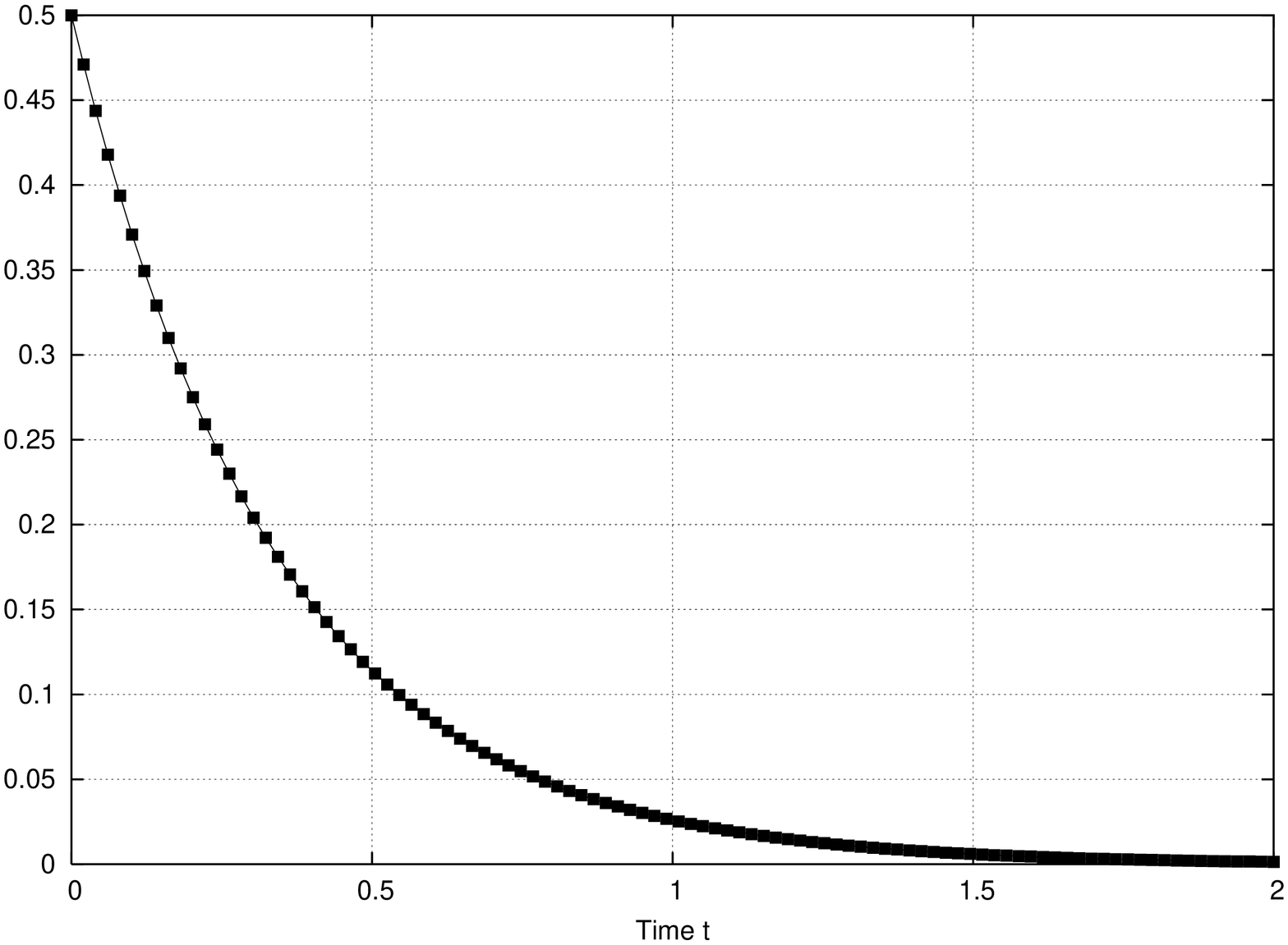}
\caption{Solid line: $\rho_x(t)$ undisturbed. Dashed line:
$\rho_x(t)$ with measurements. $\rho_x(0)=0.5$,
$\rho_y(0)=-\sqrt{1-\rho_x(0)^2}$, $\rho_z(0)=0$, $\phi=0 $, $N=1$,  $\gamma=1$
}\label{fig4}
\end{figure}

\begin{figure}[ht]
\includegraphics[scale=0.6,angle=-90]{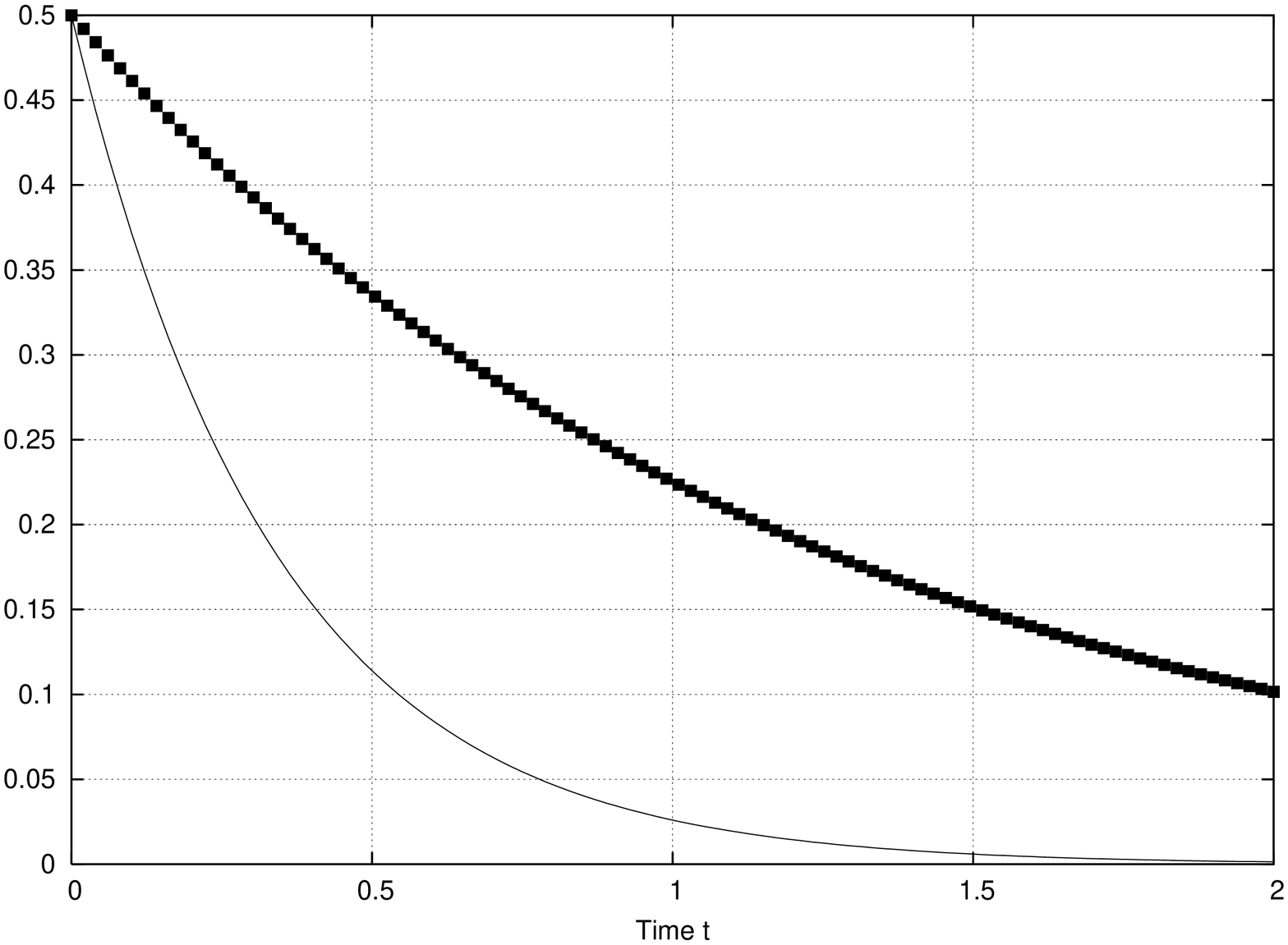}
\caption{Solid line: $\rho_x(t)$ undisturbed. Dashed line:
$\rho_x(t)$ with measurements. $\rho_x(0) =0.5$,
$\rho_y(0)=-\sqrt{1-\rho_x(0)^2}$, $\rho_z(0)=0$, $\phi=\phi_Z$, $N=1$,
$\gamma=1$ }\label{fig5}
\end{figure}

In Fig. (\ref{fig4}), it can be shown the evolution  of
$<\sigma_x>$ for $\phi=0$ with measurements and without
measurements. We observe that the evolution  is not affected  by
the measurements. This agrees with the usual assumption  that for
an unstable system with exponential decay   Zeno effect is not
observable. In Fig. (\ref{fig5}) for $\phi=\phi_Z$, one can
appreciate the reduction of the decay rate  when comparing with
the not disturbed case. For the phase $\phi=\phi_{AZ}$ the rate of
decaying grows and we have Anti-Zeno effect.

\section{Indirect measurements}

When indirect measurements are being done, the master equation
with  continuous monitoring of  $\sigma_x$ takes the form
\cite{Braginsky},
\begin{equation}
\frac{\partial \rho}{\partial t} = {\it L}(\rho) - \frac{1}{T_0} [
\, \sigma_x\, ,\, [\sigma_x,\rho]\,]
\end{equation}
where $T_0$ is the coupling constant between the measuring
apparatus and the system. Writing this equation in terms of
Bloch`s vector for the two level system in a squeezed bath  we
have,
\begin{displaymath}
\dot{\rho_x} = -\gamma \left(N+1/2+M \cos (\phi)\right) \rho_x +
\gamma M \sin(\phi) \rho_y
\end{displaymath}
\begin{displaymath}
\dot{\rho_y} = -\gamma \left(N+1/2-M \cos (\phi)\right) \rho_y +
\gamma M \sin(\phi) \rho_x -\frac{4}{T_0} \rho_y
\end{displaymath}
\begin{equation}\label{ec43}
\dot{\rho_z} = -\gamma \left(2N+1/2 \right) \rho_z
-\gamma-\frac{4}{T_0} \rho_z
\end{equation}
The limit $T_0 \rightarrow \infty$ corresponds to no measurement
being done. For  $T_0 \rightarrow  0$ equations (\ref{ec43})
transform into equations (\ref{ec39}). Then, for these kind  of
measurements one obtains similar effects that those  observed in
the previous section for the projective measurements.
\section{Conclusion}
We have presented an explicit example of a system where the
appearance of Zeno (or anti-Zeno) effect may be induced in a
regime for which it would decay exponentially if no measurements
were done. Working with a two level system in squeezed
electromagnetic bath, we found that these effects may be induced
by choosing adequately the phase of the squeezing of the bath.
This result is interpreted as the natural result of the
interaction of the equivalent spin system  with the fluctuating
fictitious magnetic field.

\section{Acknowledgments} This work was supported
by Did-Usb grant Gid-30 and by Fonacit grant G-2001000712.

\newpage

\newpage

\newpage

\newpage

\newpage

\newpage

\newpage
\begin{figure}[ht]
\includegraphics[scale=0.6,angle=-90]{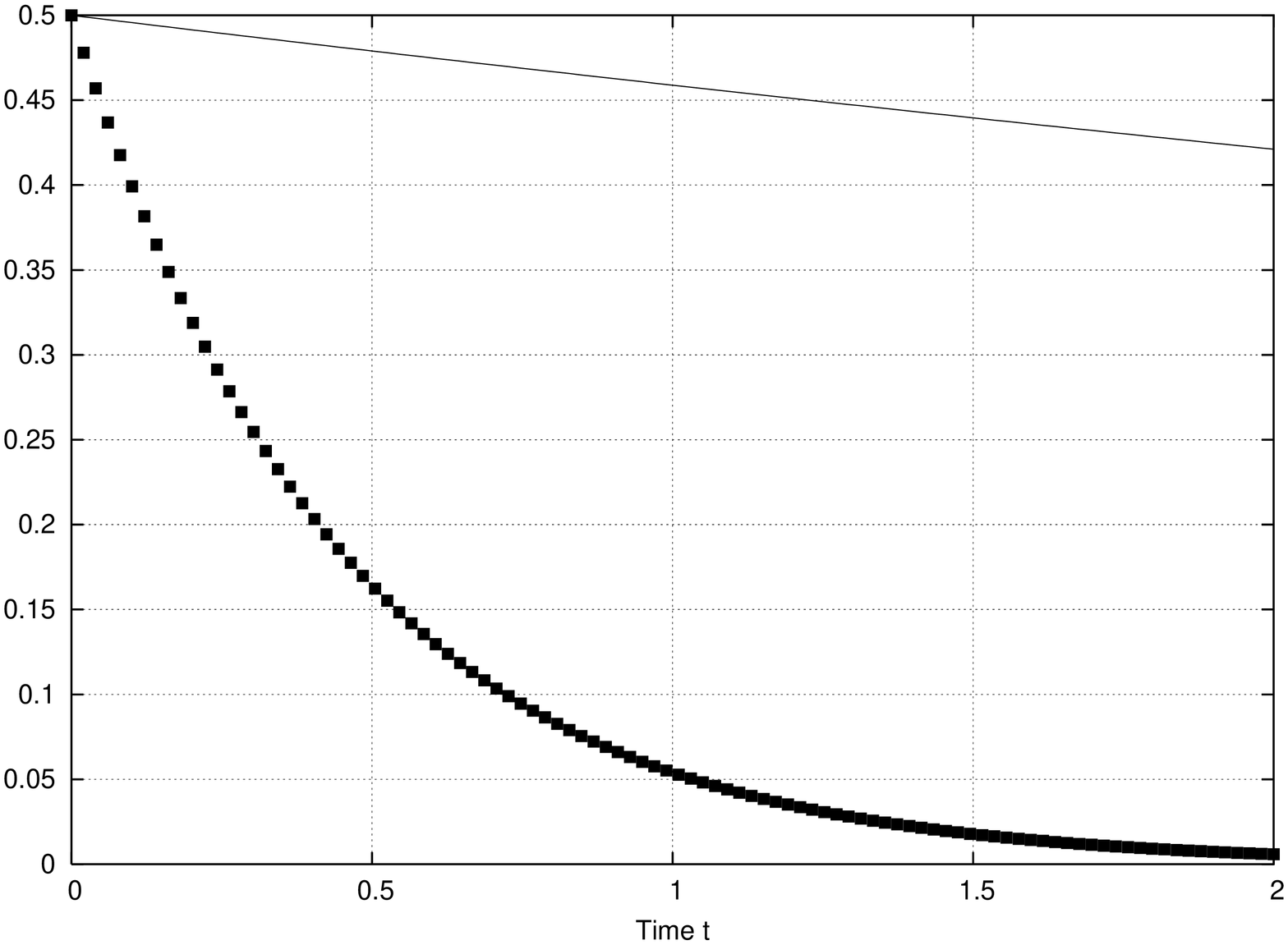}
\caption{Solid line: $\rho_x(t)$ undisturbed. Dashed line:
$\rho_x(t)$ with measurements. $\rho_x(0)=0.5$,
$\rho_y(0)=-\sqrt{1-\rho_x(0)^2}$, $\rho_z(0)=0$, $\phi=\phi_{AZ}$, $N=1$,
$\gamma=1$ }\label{fig6}
\end{figure}
\end{document}